\documentclass[aps,pra,twocolumn,showpacs,amsmath,floatfix]{revtex4}
\usepackage{graphicx}
\usepackage{float}
\usepackage{color}

\begin{document}

\hyphenpenalty=5000

\tolerance=1000

\title{Efficient Scheme of Experimental Quantifying non-Markovianity in High-Dimension Systems}

\author{S.-J. Dong}
\affiliation{Key Laboratory of Quantum Information, University of Science and Technology of China, CAS, Hefei, 230026, People's Republic of China}
\affiliation{Synergetic Innovation Center of Quantum Information and Quantum Physics, University of Science and Technology of China, Hefei, Anhui 230026, People's Republic of China}

\author{B.-H. Liu}
\affiliation{Key Laboratory of Quantum Information, University of Science and Technology of China, CAS, Hefei, 230026, People's Republic of China}
\affiliation{Synergetic Innovation Center of Quantum Information and Quantum Physics, University of Science and Technology of China, Hefei, Anhui 230026, People's Republic of China}

\author{Y.-N. Sun}
\affiliation{Key Laboratory of Quantum Information, University of Science and Technology of China, CAS, Hefei, 230026, People's Republic of China}
\affiliation{Synergetic Innovation Center of Quantum Information and Quantum Physics, University of Science and Technology of China, Hefei, Anhui 230026, People's Republic of China}

\author{Y.-J. Han}
\email{smhan@ustc.edu.cn}
\affiliation{Key Laboratory of Quantum Information, University of Science and Technology of China, CAS, Hefei, 230026, People's Republic of China}
\affiliation{Synergetic Innovation Center of Quantum Information and Quantum Physics, University of Science and Technology of China, Hefei, Anhui 230026, People's Republic of China}

\author{G.-C. Guo}
\affiliation{Key Laboratory of Quantum Information, University of Science and Technology of China, CAS, Hefei, 230026, People's Republic of China}
\affiliation{Synergetic Innovation Center of Quantum Information and Quantum Physics, University of Science and Technology of China, Hefei, Anhui 230026, People's Republic of China}

\author{Lixin He}
\email{helx@ustc.edu.cn}
\affiliation{Key Laboratory of Quantum Information, University of Science and
  Technology of China, CAS, Hefei, 230026, People's Republic of China}
\affiliation{Synergetic Innovation Center of Quantum Information and Quantum
  Physics, University of Science and Technology of China, Hefei, Anhui 230026,
  People's Republic of China}

\date{\today }

\pacs{03.65.Yz, 42.50.-p, 03.67.-a}
\begin{abstract}
The non-Markovianity is a prominent concept of the dynamics of the open
quantum systems, which is of fundamental importance in quantum mechanics and
quantum information. Despite of lots of efforts, the experimentally measuring
of non-Markovianity of an open system is still limited to very small
systems. Presently, it is still impossible to experimentally quantify the
non-Markovianity of high dimension systems with the widely used
Breuer-Laine-Piilo (BLP) trace distance measure. In this paper, we propose a
method, combining experimental measurements and numerical calculations, that
allow quantifying the non-Markovianity of a $N$ dimension system only scaled
as $N^2$, successfully avoid the exponential scaling with the dimension of the
open system in the current method. After the benchmark with a two-dimension
open system, we demonstrate the method in quantifying the non-Markovanity of a
high dimension open quantum random walk system.
\end{abstract}

\maketitle

\section{Introduction}

No real physical systems could be regarded as a purely closed system, as they
are inevitably interacting with environments. Therefore, the dynamic of the
open quantum systems is in the central of the fundamental quantum mechanics
and quantum information
science.~\cite{Breuer2007} Non-Markovianity~\cite{Angel} is the prominent
concept in open systems and attracts a lot of attention from theoretical and
experimental aspects.~\cite{Chin2012,Bylicka2013,
  Vsile2011a,Breuer2009,Laine2010,Vasile2011b,
Luo2012,Lorenzo2013} It has been shown that the non-Markovianity
can be exploited as useful resource in quantum technology. For examples: it
may be used to improve efficiency of quantum information processing and
communication;~\cite{Bylicka2013} it is benefit in quantum metrology;
~\cite{Chin2012} and it may be used to improve the security in
continuous-variable quantum key distribution,~\cite{Vasile2011} etc. It is
therefore important to quantify the non-Markovianity of a quantum
system. However, despite of the extensive investigation, characterization and
measurement of the non-Markovianity of an open system is still limited to very
small systems,~\cite{Lu2010,Xu2010,Laine2012} mostly one-qubit system in
experiments.~\cite{Liu2011}

The measurement base on trace distance, proposed by Breuer, Laine and Piilo 
(BLP),~\cite{Breuer2009} is one of the most popular definitions for
non-Markovianity, and is widely used in theoretical and experiment
investigations.~\cite{Liu2011,Liu2013}
To quantify the non-Markovianity by BLP measurement, one needs to find a pair
of initial states to maximal a function based on the trace distance. Only when
the interactions between the system and the environment are exactly known (the
dimension of the whole system is not too huge), the optimal state pair and the
measure of the non-Markovianity can be found numerically. For very limited
quantum open systems which can be exactly solved, such as the Jaynes-Cummings
model~\cite{Dalton2001} and quantum Brownian motion
model,~\cite{Intravaia2013} the measure of the non-Markovianity can be
analytically found. However, for general quantum open systems where we have no
exact information about the interaction between the system and the environment
(or the dimension of the whole system is large), the optimal initial state
pair can only be found experimentally by scanning the dynamics of the whole
initial state spaces which is a tough task even for two-dimension systems. In
a typical experiment~\cite{Liu2014} to quantify the non-Markovianity in a
two-dimension quantum open system, in order to achieve a reasonable accuracy,
total 5000 states' dynamics were measured. Even worse, the number of scanning
states grows exponentially with the dimension of the system. For a system
containing two spin-1/2 qubits (which can be viewed as a 4-dimension system),
one need to scan about $D^4$ different initial states to obtain the
non-Markovianity, where $D$ is the number of samples for each degree of
freedom. Typically, $D$ should be 100 or even more to ensure the
accuracy. Therefore even quantification the non-Markovianity of this simple
two qubits system is beyond our current experimental ability.

In this work, inspired by the idea of standard quantum process
tomography,~\cite{Chuang1997} we propose an efficient method, combining
handful experiments (polynomial scaled with the dimension
of the system) and the numerical optimization method to measure the
non-Markovianity for high dimension systems (The systems containing more than
one qubits can be regarded as high dimension systems). This method require no
prior information about the interactions between the system and its
environment. Due to the linearity of the dynamics of the quantum system, we
need only experimentally measure the dynamics of some linearly independent
states of the system, and the dynamics of the whole state space can be rebuilt
by linear combination of these experimental results. We then find the optimal
state pair in quantifying non-Markovianity through numerical calculation based
on these experiment data. The number of the measurement is scaled as $N^2$,
where $N$ is the dimension of the quantum system. Using this method, the
former intractable non-Markovianity measurement of high dimension system can
be easily investigated. After the benchmark on a two-dimension open system, we
demonstrate our algorithm to quantify the non-Markovianity of a high dimension
open quantum random walk system.

\section{Methods}

%{ \bf Simplify the measure of non-Markovianity }
The BLP measurement of the non-Markovianity~\cite{Breuer2009} is defined on the trace distance of two state $\rho_1$, $\rho_2$, that is,
$D(\rho_1,\rho_2)=\frac{1}{2}{\rm tr}|\rho_1-\rho_2|$ where $|A|=\sqrt{A^\dag A}$. If $A$ is Hermitian, ${\rm tr}|A|=\sum_i |\lambda_i|$ is the sum of the
absolute value of all the eigenvalue of matrix $A$. This quantity describe the distinguishability between the states $\rho_1$ and $\rho_2$:
if it is zero, the two states are indistinguishable, otherwise, they are
distinguishable. Based on this definition, the measure of the non-Markovianity
of an open system can be defined as,
\begin{equation}\label{eq:max_non}
\mathcal{N}={\rm max}_{\rho_{1,2}(0)}\int_{\delta>0}dt\, \delta(t,\rho_{1,2}(0))\, ,
\end{equation}
where $\delta(t,\rho_{1,2}(0))=\frac{d}{dt}D[\rho_{1}(t),\rho_{2}(t)]$ is the
change rate of the trace distance. $\rho_{i}(t)$, $i$=1, 2 is the density
matrix of the open system at time $t$ with the initial state $\rho_{i}(0)$. The time-integration is extended over all time
intervals in which $\delta$ is positive, and the maximum should be optimized
over all pairs of initial states. 
Roughly speaking, the integral intervals
stand for the time intervals when the information flows back to the system from the environment.

The most difficult task to measure the non-Markovianity is to find a 
pair of states, $\rho_1$, $\rho_2$ that maximize
Eq.~(\ref{eq:max_non}). Generally, it need experimentally scan the state pairs
in the whole parameter space. Some simplifications can be
made.~\cite{Liu2014,Wibmann2012} 
It has been rigorously proven that the optimal
states pair should be on the boundary of the physical state space and the
states pair are orthogonal each other. Therefor, we can only scan the boundary
of physical state space (For two-dimension case, these states are pure
states). Another simplification to the measure was demonstrated in
Ref. \onlinecite{Liu2014}, which illustrated that the measure can be obtained efficiently
in an arbitrary neighborhood of any fixed state in the interior of the state
space. That is, it needs only scan one state of the pair in the physical state
space. This can dramatically reduce the experimental work. However, the number
of the experiments is still too large and will exponentially increase with the
dimension of the systems.  Therefore, the non-Markovianity in higher dimension
is still intractable to quantify with the current method.
Now we introduce another scheme to simplify the experimental quantification of 
non-Markovianity in an open system which make the high dimension system reachable.

Following the idea of the quantum process tomography,~\cite{Chuang1997} the
state of a quantum open system with $N$ dimension can be expressed as
$N$$\times$$N$ density matrix. Any density matrix can be expanded by $N^2$
linear independent bases,
\begin{eqnarray}\label{rho}
   \rho^x_{mn}&=& (|m\rangle\langle n|+|n\rangle\langle m|)/2,\;\;   (m>n)
   \nonumber \\
   \rho^y_{mn}&=&i(|m\rangle\langle n|-|n\rangle\langle m|)/2,\;\;  (m>n)
   \nonumber \\
   \rho^0_m&=&|m\rangle\langle m|)\, ,
\end{eqnarray}
where $|m\rangle$ ($m=1,2,\cdots,N$) is the basis vector of the system. The
operators $\rho^x_{mn}$ ($\rho^y_{mn}$) play the similar role of the pauli
matrices $\sigma^x$ ($\sigma^y$) in two-dimension systems. Without loss of
generality, we assume that the system and the environment is in a
product state at the initial time $t=0$,
i.e. $\rho(0)=\rho_s(0)\otimes\rho_{e}(0)$, where $\rho(t)$, $\rho_s(t)$,
$\rho_{e}(t)$ are the density matrices
of the whole system (system+enviroment), the quantum system and the
environment at time $t$, respectively. Using the above introduced bases, the
state of the open system
at any time $t$ can be written as:
\begin{eqnarray}
\label{eq:rho}
 \rho_s(t)=\sum_{m>n}a^x_{mn} \rho^x_{mn}(t)+\sum_{m>n}a^y_{mn}\rho^y_{mn}(t)+\sum_m a^0_m \rho^0_m(t)\,,
\end{eqnarray}
where $a^x_{mn}$, $a^y_{mn}$, $a^0_m$ are time independent constants
determined by the initial states.
$\rho^i_{mn}(t)$ $i=x,y,0$ is the dynamics of the bases.
It suggests that the dynamics of the open system can be completely determined by the dynamics of the bases.

In experiments, the dynamics of the $N^2$ bases can be obtained by the following procedure:
\begin{enumerate}
\item  Prepare the initial states of the system to $|m\rangle$,
where $m$=1, 2, $\cdots$, $N$.
$|m\rangle$ can be any set of complete and orthogonal vector bases of the system.
By measuring the dynamics of the open system, we obtain $\rho^0_m(t)$.

\item Prepare the initial states of the system to
  $(|m\rangle+|n\rangle)/\sqrt{2}$ ($m>n$), measuring the dynamics of the open
  system. We obtain $\rho^x_{mn}(t)+\frac{1}{2}(\rho^0_m(t)+\rho^0_n(t))$.

\item Prepare the initial states of the system to
  $(|m\rangle+i|n\rangle)/\sqrt{2}$ ($m>n$), measuring the dynamics of the open
  system. We obtain $-\rho^y_{mn}(t)+\frac{1}{2}(\rho^0_m(t)+\rho^0_n(t))$.

\end{enumerate}	
We therefore have the dynamics of the $N^2$ bases of the open system. 
Using the dynamics of these bases, the BLP measure of the non-Markovianity of the open system
can be achieved by numerically optimizing the parameters, $a^x_{mn}$, $a^y_{mn}$ and $a^0_{m}$, of the initial states
by computer using Eq.~(\ref{eq:max_non}),and Eq.~(\ref{eq:rho}) through deepest descent algorithm.
The nice scale of this method make it possible to apply for the high-dimension system which is intractable for the traditional method.
It is worth noting that the basis introduced here is not unique.
Any $N^2$ such linear independent states are enough for the procedure. The
simplifications introduced in Ref.~\onlinecite{Wibmann2012} and Ref.~\onlinecite{Liu2014},
can also be used to reduce the numerical optimization efforts in this method.

\section{Results and discussion}

\subsection{Benchmark for the two-dimension system}

To demonstrate the power of our scheme, we first make benchmark tests on the
well studied two-dimension open system. In the typical
experiment,~\cite{Liu2014}
the two-dimension open quantum system is provided by the
polarization of a photon which coupled to the environment through its
frequency degree. To quantify the non-Markovianity of this system, 5000
different states in the Bloch surface have been scanned to find the optimal
pair in Eq.(1). The conclusion in \cite{Wibmann2012} has been used to simplify
the experiment,
which states that the optimal state pairs are orthogonal pure states, and
therefore, function $\mathcal{N}$ in Eq.(1) only depends on the angles between
two states on the Bloch surface.

\begin{figure}
\centering
\includegraphics[width=3.2in]{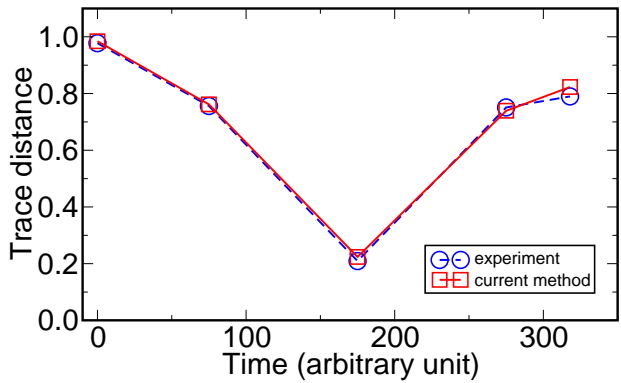}
\caption{(Color online) Comparing the trace distance directly measured from
  experiment and the results obtained from our method.
The blue circles show the trace distance as function of time
of the experimentally determined optimal state pair, by scanning 
the whole state space. The red squares show the trace distance of the optimal pair as function of time,
obtained by four basis dynamics followed by numerical optimization, as described in the main text.
}
\label{fig:compare}
\end{figure}

A $2\times2$ density matrix can be expanded by identity matrix $I$ and pauli matrixes $\sigma_x$, $\sigma_y$, $\sigma_z$ as
$\rho_s=\frac{1}{2}(I+\vec{a} \cdot \vec{\sigma})$, where vector $\vec{a}$ is
on the surface of the Bloch sphere. Using our method, 
non-Markovianity measure of this open system can be determined by the dynamics
of the pauli matrices, i.e. $\sigma_i(t)$, where $i$=$x$, $y$, $z$ which can
obtained from the dynamics of the four initial states, $|1\rangle$,
$|-1\rangle$ , $\frac{1}{\sqrt2}(|1\rangle+|-1\rangle)$ and
$\frac{1}{\sqrt2}(|1\rangle+i|-1\rangle)$ (where $|-1\rangle$ and $|1\rangle$
are two eigenvectors of the polarization photon).

Fortunately, the dynamics of these four initial states can be directly taken
from the experimental data in Ref.~\onlinecite{Liu2014}. Using these data,
we can completely determine the dynamics of any initial states using
Eq.~(\ref{eq:rho}). We then numerically find the optimal initial states pair,
and the non-Markovianity of this system. The trace distance for the optimal
initial states pair
as a function of time obtained using our scheme is compared to the directly measured one in Fig.~\ref{fig:compare}.
As we see they are in excellent agreement.
The non-Markovianity obtained from our method is 0.6, which are very close to
the measured value 0.58, both are in good agreement with the theoretical value
0.59.
The error in our method is due to the error in the measurement of the 4 basis states.

\subsection{Open quantum walk system}

\begin{figure}
\begin{center}
\includegraphics[width=3.2in]{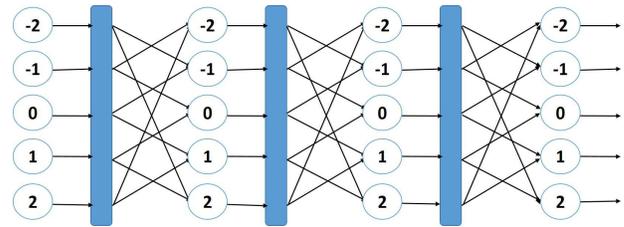}
\caption{ (color online) Cartoon of the 3-step open quantum walk system, with $X$=2.
The blue blocks denote the polarized beam splitter.
This quantum walk system includes both the lattice (location of the photon)
and the coin (polarization) which is a intrinsically high-dimension system.}
\label{QW2}
\end{center}
\end{figure}

With the confidence of the method in two-dimension open system, we apply this
method to a high-dimension open system which can not be reached
previously. Here, we demonstrate our method to qualify the non-Markovianity of
an one-dimension open quantum walk (QW) system.~\cite{ABK1,ABK2,ABK3}
which is intrinsically a high-dimension system due to the ansatz coin. QW
system is a generally interested system in quantum information, which has a
lot of application in quantum computation.
It has been shown that it is a nice tool to find new quantum algorithm and it
can be used to constitute a universal model of quantum
computation.~\cite{Childs,Venegas2012} In addition, QW has been experimentally
realized in several different
systems.~\cite{Schreiber2010,Karski2009,Peruzzo2010,Schmitz2009,Broome2010} The
open QW has attracted a lot of attention recently.~\cite{ABK1,ABK2,ABK3}

Here we study the discrete-time QW on a one-dimension lattice. The particle is
located at one site at the beginning. At each step, it can move either to the
left or to the right which is determined by the state of a coin: $|L\rangle$ (move
left) or $|R\rangle$ (move right). In quantum walk, the state of the coin can
be a superposed state. Therefore, the state of the whole QW system (including
the particle in the lattice and the coin) is
$|\psi\rangle=\sum_{x,d}C_{x,d}|x\rangle|d\rangle$ (in open QW, it should be a
density matrix) where $x=0,\pm1,\pm2,\cdots$ are the location of the particle
and $d=L,R$ are the state of the coin. The operator to make up a single step
of the QW can be defined as: $W=TC$, where $T$ is the shift operator and
defined as $T=\sum_j|j-1\rangle\langle j|\otimes| L\rangle\langle
L|+|j+1\rangle\langle j|\otimes| R\rangle\langle R|$, $C$ is the Hadamard coin
operator defined as $C=\frac{1}{\sqrt 2}(|L\rangle\langle L|+|L\rangle\langle
R|+|R\rangle\langle L|-|R\rangle\langle R|)$. The state of the system after
$N_{step}$ steps with the initial state $|\psi_0\rangle$ can be obtained as
$W^{N_{step}}|\psi_0\rangle$.

\begin{figure}
\centering
\includegraphics[width=3.2in]{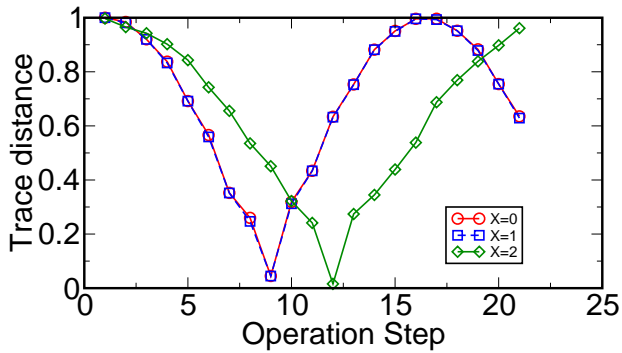}
\caption{%\textbf{The trace distance of the optimal state pairs for $X$=0, 1, 2.}.
(Color online) The trace-distances as functions of time for the optimal state pairs
obtained using our method for $X$=0 (red line), 1 (blue line) and 2 (green line).}
\label{toy}
\end{figure}

The discrete-time QW can be implemented with single photon through 
an array of beam splitters in which the coin states are mimicked by
the polarization degrees of freedom. For the open QW, the environment can be
introduced through the coupling between the frequency and the polarization
degrees of freedom
of the photon similar to the method used in Ref.~\onlinecite{Liu2014}. In this case, the
single step operator in the open QW can be modified as:
\begin{eqnarray}
	U_{step}   &=& U_{\delta t}TC, \\
	U_{\delta t} &=& \int d\omega\sum_{p=H,V}e^{in_{p}\omega\delta
          t_p}|p\rangle\langle p|\otimes|\omega\rangle\langle\omega| \, ,
\label{eq:coupling}
\end{eqnarray}
where $T$ and $C$ are shift operator and Hadamard coin operator defined
before. $U_{\delta t}$ couples the polarization($H$ or $V$) and environment to
give non-Markovianity. $n_p$ is the index of refraction for different
polarization state. $\delta t_p=\frac{L}{v_p}$ is the time that operation $TC$
takes place where $L\sim 0.5 mm$ is the thickness of the beam splitters and
$v_p$ is velocity of light with polarization $p$ in the
splitters. $|\omega\rangle$ is the environment state.
For convenience, we take the environment as a delta function , that is
$|\omega\rangle=\frac{1}{\sqrt2}[\delta(\omega-\omega_1)+\delta(\omega-\omega_2)]$. And $\omega_1=\Omega-\omega_0$, $\omega_2=\Omega+\omega_0$.
 We take the value of the parameters from the Ref.~\onlinecite{Liu2014}:
 $\omega_0=7.2\times 10^{12} s^{-1}$, $\Omega=\frac{2\pi c}{\lambda}$, where
 $\lambda$=780 nm, thus $\Omega=2.4166\times10^{15}s^{-1}$. $n_H=1.554$,
 $n_V=1.545$, $\delta t_H=Ln_H/c\sim 1.036\times 10^{-11}s$ and $\delta
 t_V\sim 1.030\times 10^{-11}s$.  In addition, we use the periodic boundary
 condition for the system, i.e. $|x_{min}-1\rangle=|x_{max}\rangle,
 |x_{max}+1\rangle=|x_{min}\rangle$ (see Fig.2), and we allow the system
 evolves for 20 steps.

\begin{table}
\caption{Comparing the number of initial states to measure 
between the present method and the direct scanning method
to obtain the non-Markovianity of the quantum walk system.
For the direct method, we assume 100 initial states to measure for each degree of freedom.
The non-Markovianity of the system is calculated using the present method.}
\begin{center}
\begin{tabular}{cccc}
	\hline\hline
      & \multicolumn{2}{c}{number of initial states} &  \\
	  X & Direc method & Present method & non-Markovianity \\
	\hline
      0 &  100$^2$         & 4    & 0.9512 \\
      1 &  100$^6$     & 36   & 0.9510 \\
      2 &  100$^{10}$  & 100  & 0.9428 \\
	\hline\hline
\end{tabular}

\label{table}
\end{center}
\end{table}

For convenience, we set the location of the particle $x$$\in$[-$X$, $X$], then the total dimension of the system is $N$=2$(2X+1)$.
To experimentally measure the non-Markovianity of this system, we need choose $N^2$ linearly independent initial states.
As introduced before, we implement experiments to get the dynamics of the following initial states:
\begin{eqnarray}\label{}
   |\psi\rangle&=&|x\rangle|d\rangle\,   \\
   |\psi\rangle&=&\frac{1}{\sqrt2}(|x_1\rangle|d_1\rangle)+|x_2\rangle|d_2\rangle),   \\
   |\psi\rangle&=&\frac{1}{\sqrt2}(|x_1\rangle|d_1\rangle)+i|x_2\rangle|d_2\rangle)
\end{eqnarray}
%w
here $x$,$x_1$,$x_2$=-$X$,$\cdots$,$X$ and $d$,$d_1$,$d_2$=$H$, $V$. With the experimental data of these initial states,
the dynamics of any initial state can be linearly constructed, and the optimal state pair can be found
by numerically searching the configuration space.

Because we do not have the experimental data at hand, we study the dynamics of
the QW system numerically using the exact diagonalization method.
We first calculate the
dynamics of the bases, then search the optimal initial states pair numerically
described in previous paragraph, and determine the non-Markovianity. We
collect the results of the trace distances of the optimal state pair for
$X$=0, 1, 2 in Fig.~\ref{toy}. The non-Markovianity for  $X$=0, 1, 2 are given
in Table \ref{table}. Actually, the result can be further confirmed via
directly exact diagonalization method in which we obtain exactly the same results.

We also compare the number of the initial states experimentally necessary to
determine the non-Markovianity of the open QW system in Table \ref{table}. The
number scales as 100$^{N}$, where $N$=4$X$+2 if previous direct scanning
method is used. Clearly it is impossible to experimentally determine the
non-Markovianity of the system for $X>$0 using this method. However, by using
our method, in which the number of initial states is only $N^2$, we are able
to measure the non-Markovianity of much larger systems.

\section{Summary}
	
We have introduced a experimental method to quantify the
non-Markovianity of high-dimension open quantum system. In our method, the
scaling of the experiment is only $N^2$ which is dramatically reduced from
exponential scaling of the conventional method. Therefore, the system which is
intractable by the former method can be easily reached with the current
method. After the benchmark with the well studied two-dimension open system,
we demonstrate the method to the high dimension open quantum walk system. This
method therefore opens up a new path to experimentally study the
non-Markovianity of high dimension open quantum system, which was impossible
previously.

The authors acknowledge the support from the Chinese National Fundamental
Research Program 2011CB921200, the National Natural Science Funds for
Distinguished Young Scholars and NSFC11374275, 11105135, 11474267, 11104261,
11374288 and the Central Universities WK2470000006, WJ2470000007.

%\bibliographystyle{apsrev}
%\bibliography{MyBib}

\end{document}